\documentclass[twocolumn,english,aps,prb,showpacs,superscriptaddress,amssymb,amsfonts]{revtex4}
\usepackage{amsmath}
\usepackage{graphicx}
\usepackage{amssymb}
\usepackage{color}
\usepackage{tabularx}
\makeatletter
\newcommand{\be}{\begin{equation}}
\newcommand{\ee}{\end{equation}}
\newcommand{\bea}{\begin{eqnarray}}
\newcommand{\eea}{\end{eqnarray}}

\newcommand{\tr}{\textrm{tr}}

\makeatother

\usepackage{babel}

\makeatother

\usepackage{babel}

\begin{document}

\title{Entanglement of Interacting Fermions in Quantum Monte Carlo}
\author{Tarun Grover}

\affiliation{Kavli Institute for Theoretical Physics, University of California, Santa Barbara, CA 93106, USA}

\begin{abstract}
Given a specific interacting quantum Hamiltonian in a general spatial dimension, can one access its entanglement properties, such as, the entanglement entropy corresponding to the ground state wavefunction? Even though progress has been made in addressing this question for interacting bosons and quantum spins, as yet there exist no corresponding methods for interacting fermions. Here we show that the entanglement structure of interacting fermionic Hamiltonians has a particularly simple form --- the interacting reduced density matrix can be written as a {\it sum} of operators that describe \textit{free fermions}.  This decomposition allows one to calculate the Renyi  entropies for Hamiltonians which can be simulated via Determinantal Quantum Monte Carlo, while employing the efficient techniques hitherto available only for  free fermions. The method presented works for the ground state, as well as for the thermally averaged reduced density matrix.
\end{abstract}


\maketitle



Quantum entanglement plays a crucial role in exposing a variety of many-body quantum  phenomena, such as topological order \cite{levin2006, kitaev2006}, surface states in quantum Hall systems and topological insulators \cite{lihaldane, lukasz}, and the universal features of critical quantum systems \cite{wilczek, cardy, ryu}. Despite its theoretical appeal, the non-local nature of the entanglement makes it a rather difficult quantity to measure in experiments, or to even evaluate numerically. In recent years, new numerical methods have been developed to calculate entanglement measures such as the Renyi entanglement entropy, for Hamiltonians of interacting bosons and quantum spin-systems \cite{hastings2010, zhang2011}. There is no obvious generalization of these techniques to fermionic systems, which differ fundamentally from bosons in the sign structure of their wavefunctions. In fact, the only known techniques for fermions are either variational in nature \cite{zhang2011, mcminis2013}, or, are restricted to one-dimensional systems \cite{white1992}, and do not address the following basic question--- \textit{given a specific Hamiltonian of interacting fermions in a general spatial dimension, how to calculate any entanglement measure in an unbiased manner?} In this article we provide an answer to this question for all Hamiltonians which can be simulated without the fermionic sign problem in the standard Determinantal Quantum Monte Carlo (DQMC) technique \cite{blank1981, white1989, assaad_rev2008}.

Our main result is in fact  more general --- we show that the reduced density matrix $\rho_A$ for an interacting fermionic system, corresponding to a subregion $A$, can be decomposed into a sum of operators that describe \textit{free fermions}. Specifically, $\rho_A = \sum_{\{s\}} P_s e^{c^{\dagger} h_s c}$ where $\{s\}$ denotes the configuration space of certain classical variables `$s$' to be introduced below, while the numbers $P_{s}$ and the matrices $h_s$ are fully determined by the underlying Hamiltonian, and we provide their general form below. This decomposition works for  the ground state, as well as for the  thermally averaged reduced density matrix at finite temperatures. Perhaps most interestingly,  it allows one to calculate highly non-local quantities, such as the Renyi entanglement entropies, in an efficient manner within DQMC, while employing the analytical techniques which were as yet available only for the free fermions 
\cite{peschel2003}. We demonstrate the method by numerically calculating the Renyi entropy $S_2$ for a one dimensional chain of 
Hubbard model, and by benchmarking it against the results from the Exact Diagonalization. Finally, we also develop a 
systematic expansion for the entanglement Hamiltonian of interacting fermions, which is again calculable 
within the Monte Carlo.

Since the notion of fermionic sign problem enters in our discussion below, we briefly mention the relevant basics \cite{hands, zhang}. Specifically, in the DQMC technique \cite{blank1981, white1989, assaad_rev2008}, the inter-particle interactions of the fermions are  re-expressed as space-time fluctuating classical fields coupled to fermion bilinears. This  allows  one to integrate out the fermions to obtain a partition function written solely in terms of the classical fields. For a class of problems, this partition function is always positive, thereby allowing one to simulate the original interacting fermionic system using the classical Monte Carlo techniques. Such problems are said to be free of the `fermion sign problem', and our results will be most useful for this same set of problems. Some of the problems that fall in this class are: Half-filled Hubbard model on bipartite lattices \cite{blank1981, white1989}, certain multi-orbital Hubbard models at any chemical potential \cite{imada, zhang}, regularized interacting Dirac fermions with an even flavors of fermions in the presence of time-reversal symmetry \cite{vafawitten,dagotto1986, hands, hands99}, $SU(2)$ gauge theory with fundamental fermions at any chemical potential \cite{dagotto1986, hands, hands99}, and $SU(N_c)$ QCD with fermions in the adjoint for any $N_c$, again at any chemical potential \cite{hands99, schafer98}.

\textit{Reduced Density Matrix in Determinantal Quantum Monte Carlo.}--- Let us recall that the full density matrix $\rho$ for a quantum Hamiltonian $H$ at a temperature $\beta^{-1}$ is given by 

\bea
\rho & = & e^{-\beta H} = \sum_i e^{-\beta E_i}|\psi_i\rangle \langle \psi_i|
\eea
 where $|\psi_i\rangle$ and $E_i$ are the eigenfunctions and eigenvalues of $H$. From this, one can define a reduced density matrix $\rho_A$ by spatially partitioning the total system into subregions $A$ and $\overline{A}$, and subsequently tracing over the Hilbert space of the subregion $\overline{A}$: $\rho_A = \tr_{\overline{A}}\, \rho$. Furthermore, one can define  entanglement measures such as the von Neumann entropy $S_{vN} = -\tr \,\,\rho_A \log(\rho_A)$ and the Renyi entropies $S_n = -\frac{1}{n-1}\log \tr(\rho^n_A)$. Our main interest lies in finding a numerically tractable expression for the reduced density matrix $\rho_A$, and the associated Renyi entropies $S_n$, for interacting fermion Hamiltonians. We find that the technique of DQMC provides a very fruitful conceptual framework to address this problem.
 
As already mentioned above, DQMC transforms a problem of interacting fermions into one of free fermions coupled to a fluctuating classical field \cite{blank1981, white1989, assaad_rev2008}. There are two different versions of this method: a zero temperature method, which is used for calculating the ground state properties, and a finite temperature method for the thermally averaged properties. For completeness, we provide an overview of these two methods in the Appendix \ref{sec:review}.  In brief, both of these schemes involve Trotter decomposition of the Hamiltonian $H$ of interest into $L_\tau$ ``time-slices'', and then introducing auxiliary classical degrees of freedom `$s$' to decouple the interacting (i.e., non-quadratic) part of the Hamiltonian. The main result of this analysis is that one can integrate out the fermions in favor of the classical fields `$s$', which are now governed by a known partition function. Returning to the original fermion problem, the expectation value of any operator $O$, either with respect to the ground state, or the thermally averaged one, may be written as \cite{blank1981, white1989, assaad_rev2008}:

\be
\langle O \rangle = \sum_{\{s\}} P_{s} \langle O \rangle _{s} \label{eq:Os}
\ee
For Hamiltonians without a sign problem, $P_{{s}}$ are positive numbers and have the interpretation of the probability distribution for the instantaneous configuration `$s$' of the classical variables, while $\langle O \rangle_{s}$ may be thought of as the expectation value of $O$ with respect to a \textit{free fermion} Hamiltonian determined also by the instantaneous configuration `$s$'. The exact form of $P_{s}$ and $\langle O \rangle _{s}$ depend on the original Hamiltonian $H$ \cite{blank1981, white1989, assaad_rev2008} and the interested reader may find explicit expressions corresponding to the Hubbard model in the Appendix \ref{sec:review}.

Perhaps most crucially, owing to the aforementioned relation to the free fermions, the expectation values $\langle O \rangle_{s}$ can be shown to follow the Wick's theorem \cite{blank1981, white1989}. For example, 
$\langle c_1^{\dagger} c_2 c_3^{\dagger} c_4 \rangle_{s} =   \langle c_1^{\dagger} c_2\rangle_{s} \langle c_3^{\dagger} c_4\rangle_{s} - \langle c_1^{\dagger} c_4\rangle_{s} \langle  c_3^{\dagger} c_2 \rangle_{s}$. As one might expect, this implies that the single particle Green's function $G_s$, with respect to a fixed configuration `$s$',  defined as $G^{ij}_{s} = \langle  c^{\dagger}_j c_i\rangle_{s}$, is sufficient to determine the expectation value $\langle O \rangle_s$ of \textit{all} operators at a fixed `$s$'.

After this brief introduction to the DQMC, we now return to our main problem, namely, the determination of reduced density matrix $\rho_A$ and associated entanglement measures, for interacting fermions. We claim that $\rho_A$ is given by the following simple expression:

\be 
\rho_A =  \sum_{\{s\}} P_{{s}}\, \rho _{A,{s}}  \label{eq:rhoA}
\ee

\noindent where

\be
\rho _{A,{s}} = C_{{s,A}} e^{-c^{\dagger} \log(G_{{s,A}}^{-1}-\mathbb{I})c} \label{eq:rhoAs}
\ee

\noindent Here the fermionic creation and annihilation operators $c,c^{\dagger}$ are restricted to region $A$, and $G_{s,A}$ is the projection of the Green's function $G_s$ to the region $A$. That is, $G^{ij}_{s,A} = G^{ij}_{s}$ for $i,j \in A$. $C_{{s,A}} = \textrm{Det}(\mathbb{I}-G_{{s,A}})$ is a normalizing coefficient that ensures  $\tr\, \rho_{A,s} = 1$. 

\textit{Proof}: $ \rho _{A,s}$ reproduces the single particle Green's function $G^{ij}_{s}$: $ \tr (\rho _{A,s} c^{\dagger}_j c_i ) = G^{ij}_{s}$ for $i,j \in A$. This  is a consequence of the fact that the reduced density matrix for a free fermionic system  \cite{peschel2003} is given by an expression identical to Eq.\ref{eq:rhoAs}, with $G_{s,A}$ replaced by the actual Green's function for the free problem \cite{footnote:herm}. Since Wick's theorem holds for a \textit{fixed configuration} `$s$', it follows that $\langle O \rangle _{s} = \tr  (\rho _{A,s}  O ) $ for all operators $O$ whose support lies in the subregion $A$. Therefore,

\bea
\tr (\rho_A O) \nonumber  & = &  \sum_{s} P_{s}\, \tr(\rho _{A,s} O) \\
& = &  \langle O \rangle
\eea
where we have used Eq.\ref{eq:Os}. Thus, the operator $\rho_A$ in Eq.\ref{eq:rhoA} reproduces the expectation value of all operators $O$ whose support lies in $A$, and therefore, indeed corresponds to the actual reduced density matrix for the region $A$ \cite{footnote:rho1rho2}.

Eqs.\ref{eq:rhoA},\ref{eq:rhoAs}, is our main result. It expresses the reduced density-matrix of an interacting fermionic system for arbitrary regions $A$, as a sum of  appropriately weighted operators that describe auxiliary free fermion systems. We emphasize that the form of $\rho_A$ in Eq.\ref{eq:rhoA} holds even for systems  which have a fermion sign problem, though in that case, not all $P_s$ will be positive, making the Monte Carlo sampling unfeasible at low temperatures.


 \begin{figure}[tb]
 \centerline{
   \includegraphics[scale = 0.27]{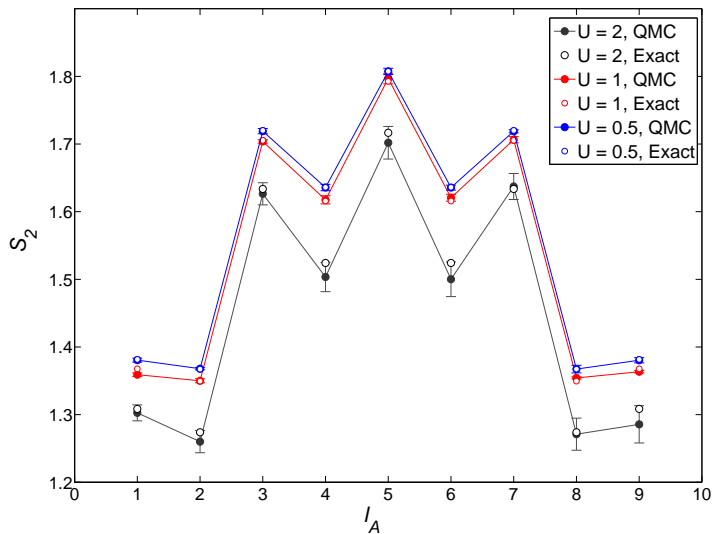}}
 \caption{Renyi entropy $S_2$ for the ground state of a 10 site single-band Hubbard model as a function of subsystem size $l_A$, for three different values of the Hubbard $U$ (with $t = 1$). Filled circles denote the DQMC results, while unfilled circles correspond to the exact results obtained via Exact Diagonalization.}
 \label{renyi10site}
 \end{figure}
 
\textit{Renyi Entropies $S_n$.}--- As a concrete application of the decomposition in Eq.\ref{eq:rhoA}, consider the Renyi entanglement entropy $S_n = -\frac{1}{n-1}\log \tr(\rho^n_A)$ for $n=2$:

\bea
S_2 & = & - \log \left[ \sum_{\{s\},\{s'\}} P_{s} P_{{s'}}\,\tr ( \rho _{A,s} \rho _{A,{s'}}) \right] \nonumber \\
& = &  - \log \left[ \sum_{\{s\},\{s'\}} P_{s} P_{{s'}} \{\textrm{Det}( G_{s} G_{{s'}}  \right. \nonumber \\ & & +  (\mathbb{I}- G_{s}) ( \mathbb{I}- G_{s'}))\} \Bigg] \label{eq:S2}
\eea
where `Det' denotes  matrix determinant. The above expression can be readily evaluated  in the Monte Carlo by sampling the expression inside the $\{\}$ brackets over two copies of the system, with the joint probability distribution function $P_{s} P_{{s'}}$. As is evident from the above expression, just the knowledge of the single particle's Green's function $G_{s}$ in the DQMC is sufficient to determine the Renyi entropy $S_2$ (or, for that matter, any Renyi entropy $S_n$ by a straightforward generalization \cite{footnote:sn}). This is rather different than the calculation of the Renyi entropies in the bosonic Monte Carlo \cite{hastings2010}, or in the Variational Monte Carlo \cite{zhang2011}, where one is required to sample a highly non-local quantity (`Swap Operator') to calculate the Renyi entropy. 

\textit{Implementation.}--- The cost of our algorithm to calculate Renyi entropies $S_n$ scales as $N^{3} L_\tau$,  where $N$ is the number of particles and $L_\tau$ is number of time slices, akin to the calculation of ground state energy or correlation functions in the DQMC \cite{blank1981, white1989, assaad_rev2008}. Indeed, one of the attractive features of our algorithm is that  it does not require any technical ingredients beyond the DQMC, because the probability distribution $P_s$, and the Green's function $G_s$ are exactly same as the ones that enter the conventional DQMC algorithm. 

We benchmarked the algorithm by  calculating $S_2$ for the ground state of a one dimensional Hubbard model:
\be 
H = H_t + H_U
\ee
\noindent where $H_t = -t\sum_{\sigma, i j} (c^{\dagger}_{i\sigma} c_{j\sigma} + c^{\dagger}_{j\sigma} c_{i\sigma}) \equiv c^{\dagger} T c$ and $H_U =  U \sum_i n_{i\uparrow} n_{i\downarrow}$. Fig.\ref{renyi10site} shows the comparison of $S_2$ obtained from a projector Monte Carlo scheme with the results from exact diagonalization for three different values of the Hubbard $U$. Clearly, the algorithm reproduces the correct result rather accurately.

\textit{Entanglement Hamiltonian.}--- The method presented allows one to also obtain an expression for the interacting `entanglement Hamiltonian' \cite{lihaldane}. The entanglement Hamiltonian is defined as $\rho_A = e^{-\mathcal{H}_A}$. For a generic interacting system, it is rather difficult to obtain a closed form expression for $\mathcal{H}_A$ as a second-quantized operator. Bosonic Monte Carlo techniques \cite{hastings2010} can only access Renyi entropies, while purely analytical techniques are as yet limited to free bosons or fermions \cite{peschel2003}, in which case $\mathcal{H}_A$ is quadratic; or include the effect of interactions via the renormalization of $\mathcal{H}_A$ that is still quadratic \cite{lauchli}.  We now show that for models that can be simulated via DQMC, a systematic expansion for $\mathcal{H}_A$ can be obtained, that includes all interactions, and  is calculable within the Monte Carlo. Let us rewrite the expression for the interacting density matrix as 

\be
\rho_A =  \sum_{\{s\}} P_{s}\, e^{-c^{\dagger}h_{s}c}
\ee
where $h_{s}$ is a matrix with components $h^{ij}_{s} = \left(\log\left(G_s^{-1} - \mathbb{I}\right)\right)^{ij}$, and we have dropped a constant shift to $h_s$. A cumulant expansion on $\rho_A$ yields $\mathcal{H}_A$:

\bea 
\mathcal{H}_A & = & - \log(\rho_A) \nonumber \\
& = & \sum_{ij} h^{ij} c^{\dagger}_i  c_j + \sum_{ijkl} k^{ijkl} c^{\dagger}_i c_j c^{\dagger}_k c_l + ... \label{eq:HA}
\eea

where $h^{ij} = \sum_{\{s\}} P_s \,\,h^{ij}_{s}$ and  $k^{ijkl} = \frac{1}{2} \left[\sum_{\{s\},\{s'\}} P_s P_{s'} \,h^{ij}_{s}  \,h^{kl}_{s'} - \sum_{\{s\}} P_s \,h^{ij}_{s} h^{kl}_{s} \right]$. One can similarly write down the higher order terms. As the above expressions show, the numbers $h^{ij}$ and $k^{ijkl}$ can be sampled within the Monte Carlo in an efficient manner, and thus, we have obtained a systematically calculable expression for the interacting entanglement Hamiltonian.



\textit{Discussion.}--- Leaving aside interactions, even the free fermions have a rather peculiar ground-state entanglement, $S \sim l^{d-1} \log l$ where $l$ is the linear extent of the entangling surface \cite{wolf2006, gioev2006}. This is in contrast to almost all other known systems where the entanglement scales as $S \sim l^{d-1}$, the so-called `area law' \cite{bombelli, srednicki}. Does a similar violation of area law holds for strongly interacting systems that do not have electron-like quasiparticles? As shown in Ref.\cite{zhang2011}, the `spinon Fermi surface' state, which is a variational wavefunction for gapless spin-liquids with Fermi surface of spinons, indeed exhibits a $l \log l$ scaling of Renyi entropy $S_2$ in two spatial dimensions, for  numerically accessible system sizes. Independently, as argued in Ref.\cite{swingle2013} on general grounds, several non-Fermi liquids can at most have a $l^{d-1} \log l$ von Neumann entanglement entropy. In the light of these results, models such as the one studied in Ref.\cite{sachdev2012} using DQMC, provide a unique opportunity to explore entanglement scaling via the method presented here, since in this model one can access fermionic quantum criticality in the presence of a Fermi surface without encountering a sign problem. On this note, it is worth mentioning that there is a large class of problems which do not have a fermion sign problem even at a finite density of fermions, including multi-orbital Hubbard models \cite{imada, hands, zhang}. These models provide a platform to boost our current understanding of entanglement in the Fermi liquids \cite{swinglefermi, mcminis2013}.

The method presented also allows one to detect `topological order' via entanglement \cite{levin2006, kitaev2006} in models that can be simulated via DQMC. For example, it was recently suggested  \cite{meng}  that the Hubbard model on the Honeycomb lattice exhibits a $\mathbb{Z}_2$ spin-liquid at intermediate $U/t$. Later, in a different study \cite{sorella, assaad}, it was argued that the aforementioned conclusion about the presence of topological order is incorrect, and the system instead exhibits a direct phase transition from a semi-metal to an anti-ferromagnet. It will be  interesting to revisit this problem via the algorithm presented here, which can potentially confirm or rule out the presence of topological order. The method is also directly applicable to several lattice matter-gauge theories \cite{hands, dagotto1986, hands99}, and could be useful, for example, in exploring the possibility of topological superconductivity in $SU(2)$ matter-gauge theory \cite{dagotto1986, hands99, nishida}. 
One can also study entanglement in sign problem free spin-systems, such as the spin-1/2 Heisenberg model on the square lattice, by studying the large $U$ limit of the Hubbard model at half-filling. This will provide an alternate viewpoint as compared to the bosonic Monte Carlo \cite{hastings2010}. 

Finally, we note that the expansion in Eq.\ref{eq:HA} for $\mathcal{H}_A$ might help in understanding which specific interacting systems have a (non)-local entanglement Hamiltonian. The problem of determining the locality of $\mathcal{H}_A$ for a given problem has been essentially reduced to understanding the locality of the average $\langle h_s \rangle$ with respect to the probability distribution $P_s$. The expansion for $\mathcal{H}_A$ is also suggestive of an area law scaling for the entanglement entropy  $S \sim l^{d-1}$, upto multiplicative logarithmic corrections \cite{wolf2006, gioev2006}, for the ground states of $d$ dimensional systems which can be simulated without a sign problem \cite{footnote:bound}. This is because the contribution of the first term in the expansion, $\langle h_s \rangle$ , to the entanglement entropy is likely to be an area law, again upto multiplicative logarithmic corrections, because the individual terms $h_{s}$ themselves correspond to free fermion problems, and the averaging converges due to the lack of a sign problem. This would imply that $h$ behaves essentially as a $d-1$ dimensional system, and thus the contribution to the entanglement entropy from higher order terms in the expansion can be expected to scale in a similar fashion. This argument is by no means rigorous, and we leave such explorations for the future.

To summarize, we showed that the reduced density matrix for interacting fermions can be expressed rather simply in terms of free fermions, and we used this fact to develop an algorithm to calculate  the Renyi entanglement  entropies, and the entanglement Hamiltonian, for interacting fermionic systems in general dimensions. Our method provides an information-theoretic reformulation of the DQMC method, while opening a pathway to  explore many-body entanglement in strongly correlated fermionic systems.

\underline{\textbf{Acknowledgments:}} I thank Matthew Fisher and Max Metlitski for  useful discussions, and Brian Swingle, Bryan Clark and Madhav Mani for helpful comments on the draft. This research was supported in part by the National Science Foundation under Grant No. NSF PHY11-25915.

\appendix

\section{Overview of Determinantal Quantum Monte Carlo} \label{sec:review}

For completeness, here we provide a lightening overview of the DQMC scheme \cite{blank1981, white1989, assaad_rev2008}. We closely follow the notation in the review article by F. Assaad and H. Evertz \cite{assaad_rev2008}.

\subsection{Zero Temperature Projector Algorithm} \label{sec:zeroT}
In the projector scheme one represents the ground state $|\Psi\rangle$ of the interacting Hamiltonian $H$ as:

\be 
|\Psi\rangle = \lim_{\Theta \to \infty} e^{-\Theta H} |\Psi_T\rangle \label{wfnproj}
\ee
where $|\Psi_T\rangle$ is a ``trial state'' and it is assumed that $\langle\Psi_T|\Psi\rangle \neq 0$.  For practical implementation, one chooses $\Theta$ to be finite and large.
%
For concreteness, we will review the method for the Hubbard model
\be 
H = H_t + H_U
\ee
\noindent where $H_t = -t\sum_{\sigma, i j} (c^{\dagger}_{i\sigma} c_{j\sigma} + c^{\dagger}_{j\sigma} c_{i\sigma}) \equiv c^{\dagger} T c$ and $H_U =  U \sum_i n_{i\uparrow} n_{i\downarrow}$. We denote total number of particles by $N_p$ and the number of sites by $N_s$. One approximates $e^{-2\Theta H} $ as $e^{-2\Theta H} = (e^{-\Delta_\tau H_U} e^{-\Delta_\tau H_t})^{L_\tau} + O(\Delta^2_\tau)$, where $\Delta_\tau L_\tau = 2\Theta$. Decoupling the quartic fermion interaction in $H_U$ via

\bea
e^{-\Delta_\tau H_U} & = & C \sum_{\{s_i = \pm 1 \}} e^{\alpha \sum_i s_i(n_{i\uparrow}-n_{i\downarrow})} \nonumber \\
& \equiv & C \sum_{\{s_i = \pm 1 \}} e^{c^{\dagger} V(s)c}
\eea

\noindent where $\cosh(\alpha) = e^{\frac{U\Delta_\tau }{2}}$ and $C$ is a constant, one leads to a problem of fermions coupled to Ising spins $\{s_i\}$ that live on each lattice site of $m$ different ``time-slices''. Taking the trial wavefunction $|\psi_T\rangle$ as a single Slater determinant: $|\Psi_T \rangle = \prod_{i=1}^{N_p} \left(\sum_x c_x^{\dagger}  P_{x,y} \right) |0\rangle$, where $P$ is a rectangular matrix of size $N_s \times N_p$. Within this scheme, the normalization factor $\langle \Psi|\Psi\rangle$ can be thought of as a partition function of an equivalent bosonic problem, and can be shown to be equal to

\be 
\langle\Psi|\Psi\rangle = C^{L_\tau} \sum_{\{s\}} \textrm{Det} \left[ P^{\dagger} B_s\left( 2\Theta,0\right) P \right]
\ee

where $B_s$ can be thought of as discrete imaginary time propagator, $B_s(\tau_2, \tau_1) = \prod_{n = n_1 + 1}^{n_2} e^{V(\{s_n\})} e^{-\Delta_\tau T}$.

The expectation value of an operator $O$ in the ground state is given by

\bea 
\langle O \rangle & = &  \frac{\langle \Psi_T| e^{-\Theta H} O e^{-\Theta H} |\Psi_T\rangle} {\langle \Psi_T| e^{-2\Theta H}|\Psi_T\rangle} \\
& = & \sum_{\{s\}} P_{s} \langle O \rangle _{s} \label{eq:op_avg}
\eea

\noindent where

\be 
P_{s} = \frac{ \textrm{Det} \left[ P^{\dagger} B_s\left( 2\Theta,0\right) P \right]}{\sum_{\{s\}}  \textrm{Det} \left[ P^{\dagger} B_s\left( 2\Theta,0\right) P \right]} \label{eq:OszeroT}
\ee

and 

\be 
 \langle \hat{O}\rangle _{s} = \frac{\langle\Psi_T |U_s(2\Theta,\Theta) O U_s(\Theta,0)|\Psi_T\rangle}{\Psi_T|U_s(2\Theta,0)|\Psi_T\rangle}  \label{eq:PszeroT}
\ee

with $U_s$ is the operator analog of $B_s$: $U_s(\tau_2, \tau_1) = \prod_{n=n_1 + 1}^{n_2}  e^{c^{\dagger}V(\{s_n\}) c} e^{-c^{\dagger}\Delta_\tau T c}$.

\subsection{Finite-temperature Algorithm} \label{sec:finiteT}

In the finite temperature algorithm, one directly works with the partition function at inverse temperature $\beta$. Once again, for illustration, we restrict our discussion to the Hubbard model:

\bea
Z & = & \tr \,\,e^{-\beta H} \\
& = & \tr  \left[(e^{-\Delta_\tau H_U} e^{-\Delta_\tau H_t})^{L_\tau}\right] + O(\Delta^2_\tau)
\eea

Similar to the case of zero temperature algorithm, one now decouples the quartic fermionic interaction via classical Ising variables $s$ and obtains an expression similar to Eqn.\ref{eq:op_avg}. Specifically, 

\bea
\langle O \rangle & = &  \frac{\tr \left( e^{-\beta H} O \right)}{\tr \left( e^{-\beta H}  \right)} \nonumber \\
& = &  \sum_{\{s\}} P_{s} \langle O \rangle _{s}
\eea

where 

\be 
P_{s} = \frac{ \textrm{Det} \left(1 + B_s\left( \beta,0\right)\right) }{\sum_{\{s\}}  \textrm{Det} \left(1 + B_s\left( \beta,0\right)\right) } \label{eq:PsfiniteT}
\ee

and 

\be 
 \langle \hat{O}\rangle _{s} = \frac{\tr \left[U_s(\beta,\beta/2) O U_s(\beta/2,0) \right]}{ \tr \left[ U_s(\beta,0) \right]} \label{eq:OsfiniteT}
\ee

\noindent where $B_s$ and $U_s$ have the same functional form as in the zero temperature algorithm, with the replacement $2\theta \rightarrow \beta$.

\section{Correlation functions with respect to $\rho_{A,s}$} \label{sec:corr}

Consider the expression in the main text for  $\rho_{A,s}$:

\be
\rho _{A,s} = C_{s,A} e^{-c^{\dagger} \log(G_{s,A}^{-1}-\mathbb{I})c}
\ee
where  $C_{s,A} = \textrm{Det}(\mathbb{I}-G_{s,A})$. Even though the matrix $\rho_{A,s}$ is generically not Hermitian, the above expression reproduces correlation functions of all operators whose support lies in the region $A$, for a fixed configuration of classical fields $s$. First, consider the two-point correlator $\langle c^{\dagger}_1 c_2 \rangle_{s}$ for any two sites 1,2 $\in A$ :

\bea
\langle c^{\dagger}_1 c_2 \rangle_{s} & = & \tr \,\,\left(\rho_{A,s} c^{\dagger}_1 c_2 \right)  \nonumber \\ 
& = & \frac{\partial}{\partial \eta} \tr \left[ C_{s,A} e^{-c^{\dagger} \log(G_{s,A}^{-1}-\mathbb{I})c} e^{\eta c^{\dagger}M c}\right]\bigg|_{\eta = 0}
\eea
where $M^{ij} = \delta_{i,1} \delta_{j,2} $. Performing the trace, one obtains,

\bea
\langle c^{\dagger}_1 c_2 \rangle_{s} & = & \frac{\partial}{\partial \eta} \textrm{Det} (\mathbb{I} + \eta G_{s,A} M)\bigg|_{\eta = 0} \nonumber \\
& = & G^{21}_{s,A} \nonumber \\
& = & G^{21}_{s}
\eea
as expected. Let us now look at the higher point correlators. For example, consider

\bea 
\langle c^{\dagger}_1 c_2 c^{\dagger}_3 c_4\rangle_{s} & = &  \tr \,\,\left(\rho_{A,s} c^{\dagger}_1 c_2 c^{\dagger}_3 c_4 \right) \nonumber \\ & = & \frac{\partial^2 \tr \left[ C_{s,A} e^{-c^{\dagger}  \log(G_{s,A}^{-1}-\mathbb{I})c} e^{\eta_1 c^{\dagger}M_1 c} e^{\eta_2 c^{\dagger}M_2 c}\right]\bigg|_{\eta_1,\eta_2 = 0}}{\partial \eta_1 \eta_2} \nonumber 
\eea
where $M^{ij}_1 = \delta_{i,1} \delta_{j,2} $ and  $M^{ij}_2 = \delta_{i,3} \delta_{j,4}$. Performing the trace,

\bea
\langle c^{\dagger}_1 c_2 c^{\dagger}_3 c_4\rangle_{s}  & = &  \frac{\partial^2 \textrm{Det}(\mathbb{I} + \eta_1 G_{s,A} M_1 + \eta_2 G_{s,A} M_2) \bigg|_{\eta_1,\eta_2 = 0}}{\partial \eta_1 \eta_2} \nonumber \\
& = & G^{21}_{s,A} G^{43}_{s,A} - G^{23}_{s,A} G^{41}_{s,A} \nonumber \\
& = & G^{21}_{s} G^{43}_{s} - G^{23}_{s} G^{41}_{s}
\eea
That is, $\rho_{A,s}$ again reproduces the correct correlation function, as determined by the Wick's theorem. One can similarly generalize the above calculation to any operator $O$ and confirm that $\langle O \rangle _{s} = \tr  (\rho _{A,s}  O ) $.

\begin{thebibliography}{x}
\bibitem{kitaev2006} A. Kitaev, J. Preskill,  Phys. Rev. Lett. 96, 110404 (2006).

\bibitem{levin2006} M. Levin, X.-G. Wen, Phys. Rev. Lett. 96, 110405 (2006).

\bibitem{lihaldane} H. Li, F. D. M. Haldane, Phys. Rev. Lett., 101 (2008) 010504.

\bibitem{lukasz} L.  Fidkowski, Phys. Rev. Lett. 104, 130502 (2010).
\bibitem{wilczek} C. Holzhey, F. Larsen, F. Wilczek, Nucl. Phys. B 424, 443 (1994). 

\bibitem{cardy} P. Calabrese and J. Cardy, J. Stat. Mech. (2004) P06002.

\bibitem{ryu} S. Ryu, T. Takayanagi, Phys. Rev. Lett. 96, 181602 (2006).

\bibitem{hastings2010} M. B. Hastings, I. Gonzalez, A. B. Kallin, R. G. Melko, Phys. Rev. Lett. 104,
157201(2010).

\bibitem{zhang2011} Y. Zhang, T. Grover, A. Vishwanath, Phys. Rev. Lett. 107, 067202 (2011).

\bibitem{mcminis2013} J. McMinis, N. M. Tubman, Phys. Rev. B 87, 081108(R) (2013).

\bibitem{white1992} S. R. White, Phys. Rev. Lett. 69, 2863 (1992).

\bibitem{blank1981} R. Blankenbecler, D. J. Scalapino, and R. L. Sugar, Phys. Rev. D 24, 2278 (1981).

\bibitem{white1989} S. R. White, D. J. Scalapino, R. L. Sugar, E. Y. Loh, J. E. Gubernatis, and
R. T. Scalettar, Phys. Rev. B 40, 506 (1989).

\bibitem{assaad_rev2008} F. F. Assaad and H. G. Evertz, Worldline and Determinantal Quantum Monte Carlo Methods for Spins, Phonons and Electrons, in \textit{Computational Many-Particle Physics}, H. Fehske, R. Shnieider, and A. Weise Eds., Springer, Berlin (2008).



\bibitem{zhang} C. Wu, S.-C. Zhang, Phys. Rev. B 71, 155115 (2005).

\bibitem{imada} Y. Motome and M. Imada, J. Phys. Soc. Jpn. 66, 1872 (1997).

\bibitem{vafawitten} C. Vafa and E. Witten, Phys. Rev. Lett. 53, 535 (1984).

\bibitem{dagotto1986} E. Dagotto, F. Karsch, and A. Moreo, Phys. Lett. B 169,  421 (1986); E. Dagotto, A. Moreo, and U. Wolff, Phys. Rev. Lett. 57,  1292 (1986).

\bibitem{hands} S. Hands, I. Montvay, S. E. Morrison, M. Oevers, L. Scorzato,
and J. Skellurd, Eur. Phys. J. C 17, 285 (2000).

\bibitem{hands99} S. Hands, J.B. Kogut, M.-P. Lombardo, S.E. Morrison, Nucl. Phys. B 558, 327 (1999).

\bibitem{schafer98} T. Sch\"{a}fer, Phys. Rev. D 57, 3950 (1998).

\bibitem{peschel2003} I.  Peschel,  J. Phys. A: Math. Gen. 36, L205 (2003).


\bibitem{footnote:herm} The only technical difference in our case is that the matrices $G_{s}$ are not necessarily Hermitian, i.e., $G^{ij}_s \neq (G^{ji}_s)^{*}$. This is  because $G_s$ can be thought of as an equal time Green's function for a free fermion problem in the presence of a \textit{time-dependent} classical field `$s$'. Wick's theorem of course continues to hold for $\rho_{A,s}$, which is all we require (see Appendix \ref{sec:corr} for details).

\bibitem{footnote:rho1rho2} Note that the reduced density matrix is \textit{uniquely} determined by the condition that $\langle O \rangle = \tr \left(\rho_A O\right)$ for all operators $O$. Indeed, if $\tr \left(\rho_A O\right) = \tr \left(\tilde{\rho}_A O\right)$ for all $O$, then one prove $\rho^{\alpha \beta}_{A} = \tilde{\rho}^{\alpha \beta}_A$ element-wise by choosing appropriate $O$ that pick-out a particular element ($\alpha, \beta$) of $\rho_A, \tilde{\rho}_A$.


\bibitem{footnote:sn} Explicitly, Renyi entropy $S_n$ in the Monte Carlo is given by $S_n = -\frac{1}{n-1}\log \tr(\rho^n_A)$ where 
\be 
\tr (\rho^n_A) = \sum_{\{s_\alpha\}} \prod_{\beta=1}^{n} P_{s_\beta} \,F_A(s_1,s_2,...,s_n) 
\ee

and 

\be 
F_A(s_1,s_2,...,s_n) = \left(\prod_{\beta=1}^{n} C_{s_\beta,A}\right)\textrm{Det}\left(\mathbb{I} +  \prod_{\gamma=1}^{n} (G^{-1}_{s_\gamma,A}-\mathbb{I})^{-1}\right) \nonumber
\ee

Thus, one samples the quantity $F_A(s_1,s_2,...,s_n)$ with the probability distribution $\prod_{\beta=1}^{n} P_{s_\beta}$ over $n$ copies of the system.

\bibitem{lauchli} V. Alba, M. Haque, A. M. L\"{a}uchli, Phys. Rev. Lett. 110, 260403 (2013).

\bibitem{wolf2006} M. M. Wolf, Phys. Rev. Lett. 96, 010404 (2006).

\bibitem{gioev2006} D. Gioev and I. Klich, Phys. Rev. Lett. 96, 100503 (2006).



\bibitem{bombelli} L. Bombelli, R. K. Koul, J. Lee, R. D. Sorkin, Phys. Rev. D 34, 373 (1986).

\bibitem{srednicki} M. Srednicki, Phys. Rev. Lett. 71, 66 (1993).

\bibitem{swingle2013} B. Swingle, T. Senthil, Phys. Rev. B 87, 045123 (2013).

\bibitem{sachdev2012} E. Berg, M. A. Metlitski, S. Sachdev, Science 338, 1606 (2012).

\bibitem{swinglefermi} B. Swingle, Phys. Rev. B 86, 045109 (2012); B. Swingle, arXiv:1209.0769.

\bibitem{meng} Z. Y. Meng, T. C. Lang, S. Wessel, F. F. Assaad, A. Muramatsu, Nature 464, 847 (2010).

\bibitem{sorella} S. Sorella, Y. Otsuka, S. Yunoki, Sci. Rep. 2, (2012).

\bibitem{assaad} F. F. Assaad, I. F. Herbut, arXiv:1304.6340.

\bibitem{nishida} Y. Nishida, Phys. Rev. D 81, 074004 (2010).

\bibitem{footnote:bound} One might have guessed that the expression in  Eq.\ref{eq:rhoA} provides useful upper and lower bounds on the von Neumann entanglement entropy by employing the concavity of the density matrices. However, the matrices $\rho_s$ are not Hermitian, which invalidates such an argument.

\end{thebibliography}
\end{document}